\documentclass[reqno]{amsart}
\usepackage{amssymb,graphics,graphicx,bbm,hyperref,color, float}

\definecolor{gray}{rgb}{0.9,0.9,0.9}

\def\be{\begin{equation}}
\def\ee{\end{equation}}
\def\bm{\begin{multline}}
\def\bfig{\begin{figure}[htb]}
\def\efig{\end{figure}}
\newcommand{\dd}{{\rm d}}
\newcommand{\e}[1]{\,{\rm e}^{#1}\,}
\newcommand{\ii}{{\rm i}}

\numberwithin{equation}{section}
\newtheorem{theorem}{Theorem}[section]

\newtheorem{lemma}[theorem]{Lemma}

\newtheorem{assumption}{Assumption}

\newcommand{\eps}{{\varepsilon}}
\newcommand{\bbC}{{\mathbb C}}
\newcommand{\bbN}{{\mathbb N}}
\newcommand{\bbR}{{\mathbb R}}
\newcommand{\bbX}{{\mathbb X}}
\newcommand{\bsg}{{\boldsymbol g}}
\newcommand{\bsk}{{\boldsymbol k}}

\newcommand{\bsn}{{\boldsymbol n}}

\newcommand{\bsz}{{\boldsymbol z}}
\newcommand{\bsrho}{{\boldsymbol\rho}}
\newcommand{\caB}{{\mathcal B}}
\newcommand{\caC}{{\mathcal C}}
\newcommand{\caE}{{\mathcal E}}

\newcommand{\caS}{{\mathcal S}}
\newcommand{\caV}{{\mathcal V}}

\definecolor{orange}{rgb}{0.9,0.4,0}

\makeatletter
  \def\tagform@#1{\maketag@@@{\tiny{(#1)}\@@italiccorr}}
\makeatother
\renewcommand{\eqref}[1]{(\ref{#1})}

\begin{document}


\title{Multispecies Virial Expansions}

\author[S. Jansen]{Sabine Jansen}
\address{Leiden University, Postbus 9512,
      2300 RA Leiden, The Netherlands}
\email{sabine.jansen@math.leidenuniv.nl}

\author[S.J. Tate]{Stephen J. Tate}
\address{Department of Mathematics, University of Warwick,
Coventry, CV4 7AL, United Kingdom}
\email{s.j.tate@warwick.ac.uk}

\author[D.Tsagkarogiannis]{Dimitrios Tsagkarogiannis}
\address{Department of Applied Mathematics, University of Crete, P.O. Box 2208, 71003, Heraklion, Greece}
\email{tsagkaro@tem.uoc.gr}

\author[D. Ueltschi]{Daniel Ueltschi}
\address{Department of Mathematics, University of Warwick,
Coventry, CV4 7AL, United Kingdom}
\email{daniel@ueltschi.org}



\begin{abstract}
We study the virial expansion of mixtures of countably many different types of particles. The main tool is the Lagrange-Good inversion formula, which has other applications such as counting coloured trees or studying probability generating functions in multi-type branching processes. We prove that the virial expansion converges absolutely in a domain of small densities. In addition, we establish that the virial coefficients can be expressed in terms of two-connected graphs.
\end{abstract}

\thanks{\copyright{} 2013 by the authors. This paper may be reproduced, in its
entirety, for non-commercial purposes.}

\maketitle

\section{Introduction}
\label{sec intro}

One of the central questions of statistical mechanics
is the calculation of the thermodynamic properties
of a given fluid 
starting 
from the forces between the molecules.
A major contribution in this direction is the work of Mayer and his collaborators \cite{M37,
MA, HM, MH}. This was further developed by Born and Fuchs \cite{BF}, and Uhlenbeck and Kahn \cite{UK}.
All of these papers concern the theory of
non-ideal gases for which the pressure is written as a power series in the activity or density.
The convergence of such power series was investigated much later, during the 1960's, 
by Groeneveld, Lebowitz, Penrose and Ruelle.
Parallel to the study of monoatomic gases, multi-component systems were also investigated, for example in \cite{M39} for the case of a two-component system
and later in \cite{F42} for a mixture with an arbitrary (but finite) number of different components.
Although briefly mentioned already in \cite{F42},
the complete study of the convergence of the activity expansion comes later in \cite{BL} 
for mixtures of finitely many components.

The generalisation of virial expansions and approximate equations of state 
from the monoatomic gas to mixtures is not straightforward. 
The van der Waals equation for binary mixtures of hard spheres, for example, comes in different versions, distinguished by different ``mixing rules'' for obtaining effective parameters of the mixture~\cite{henderson-leonard70}; see also~\cite{LR} for the Percus-Yevick and virial equations of state. 
Going from binary mixtures to multi-component systems is not easy either -- in Fuchs' words, ``for the most intricate parts of the calculation even the theory of the two-component system does not give any indication as to the results for the many-component system (e.g. the reduction of the cluster integrals to the irreducible cluster integrals)''~\cite{F42}.
Even less trivial is the extension from finitely to infinitely many types of particles, in particular for estimating the domain of convergence.

The present work deals with the  virial expansion for countably many types of particles, addressing in particular the problem of convergence.  
Our motivation is two-fold. First, mixtures as studied in physical chemistry are of interest on their own. 
Second, multi-component systems arise as effective models as in ``the consideration of certain phenomena connected
with the order-disorder transition in alloys" \cite{F42}, or in
the treatment of a monoatomic gas as a mixture of ``droplets''  (groups of particles close in space) \cite[Chapter 5.27]{hillbook}. A droplet can comprise arbitrarily many particles, and there can be arbitrarily many droplet sizes -- this is why we are interested in mixtures with infinitely many species.

In the first part of this article, we are given a pressure function that depends on countably many fugacity parameters. This yields countably many densities (that are also functions of fugacities) and the goal is to write the pressure as function of densities. In the case of one parameter, this can be done using Lagrange inversion~\cite{lebowitz-penrose64}. We note that this analytical method is also a standard tool in combinatorics~\cite[Chapter 3]{BLL}. In the case of multicomponent systems, the key ingredient is Good's generalisation of the Lagrange inversion to several variables~\cite{good60}. The Lagrange-Good inversion has attracted attention in a variety of contexts ~\cite{abdesselam03,bruijn83,Ges,ehrenborg-mendez94}. Faris has recently noticed its relevance for the virial expansion \cite{Far12}. Our main focus is on the convergence of the expansion. Despite recent renewed interest on the single-species virial expansion~\cite{Jan,PT,tate13,morais-procacci13}, this question does not seem to have been addressed before in the case of infinitely many species. Our main result is the existence of a non-trivial domain of convergence for the expansion of the pressure as function of densities. A novel feature is to use Lagrange-Good for proving convergence.

The second part of this article deals with a gas of classical particles with two-body interactions. There are many species of particles. Under some assumptions, the virial coefficients are given in terms of two-connected graphs (irreducible cluster integrals). Early derivations for systems with one, two, or finitely many components can be found in~\cite{BF,M39,F42}. 
Thanks to the work of Mayer, a systematic connection with combinatorics has been initiated, linking the enumeration of connected and two-connected graphs with cluster and virial expansions. This connection 
has been further developed in the
work of Leroux and collaborators \cite{L}, leading to modern proofs for the expression of virial coefficients. This was generalised by Faris to the case of many species \cite{Far12}. The present article has some overlap, but there are several differences. In particular, we formulate sufficient conditions on the interactions that guarantee the convergence of the virial expansion.

We would like to emphasise that our results have relevance beyond statistical mechanics. Indeed, our main result can be formulated as an inverse function theorem for functions between Banach spaces that are not necessarily Fr{\'e}chet-differentiable (Section~\ref{sec:inverse-function}). In addition, our result can be applied in the original context of the Lagrange-Good inversion formula: Good motivated his work by stochastic branching processes and combinatorics of coloured trees \cite{good60,good65}. Recursive properties of trees lead to functional equations between generating functions. When combined with the inversion formula they yield expressions of probabilities or tree cardinalities as contour integrals; Good explicitly computed some of those integrals. Our result yields  bounds for cardinalities without having to compute the integrals, which may come in handy when explicit computations prove too complicated.

The article is organised as follows. In Section \ref{sec virial} we give the general setting of the virial expansion
in the context of formal power series. The main theorem proposes sufficient conditions under which the virial
power series is absolutely convergent. This is done by deriving explicit bounds on the virial coefficients
via the Lagrange-Good inversion formula.
In Section \ref{sec graphs} we prove a dissymmetry theorem for coloured weighted graphs and deduce that the virial coefficients are given by the two-connected
graphs; this result holds whenever the weights satisfy a block factorisation property.
Finally, in Section \ref{sec clexp}, we consider a mixture of rigid molecules and, using the results on the convergence of the cluster expansion given in \cite{Uel,PU}, we show that the mixture meets the conditions of Section~\ref{sec virial}.

\section{General virial expansions}
\label{sec virial}

\subsection{Setting \& results}

Let $\bsz = (z_{1}, z_{2}, \dots)$ denote a sequence of complex numbers. Consider the formal series
\be
\label{p(z)}
p(\bsz) = \sum_{\bsn} b(\bsn) \bsz^{\bsn},
\ee
where the sum is over all multi-indices $\bsn = (n_{1}, n_{2}, \dots)$, $n_{i} \in \bbN$, with finitely-many non-zero entries
and $\sum_{i\in \bbN} z_i \geq 1$. We assume that for all $i \in \bbN$, the coefficient $b(0,\ldots,1,0,\ldots)$ of $z_i$ in $p(\bsz)$ is non-zero; this is the only condition needed for Lagrange inversion techniques. In Sections~\ref{sec graphs} and~\ref{sec clexp} we will need the additional assumption that the coefficients are normalized as $b(0,\ldots,1,0,\ldots)=1$, as is the case in most applications in Statistical Mechanics. 

We use the notation $\bsz^{\bsn} = \prod_{i} z_{i}^{n_{i}}$.  In statistical physics, $z_{i}$ represents the ``activity'' (or ``fugacity'') of the species $i$, and $p(\bsz)$ is the pressure of the system with many species. A physically relevant quantity is the density $\rho_{i}$ of the species $i$, whose definition is
\be
\label{def densities}
\rho_{i}(\bsz) = z_{i} \frac{\partial p}{\partial z_{i}}(\bsz).
\ee
We do not suppose yet that the series for $p(\bsz)$ is convergent, and the equation above should be understood in the sense of formal series. To be precise, $\rho_{i}(\bsz)$ is the formal series with coefficients $n_{i} b(\bsn)$.

In statistical physics, the {\it virial expansion} is the expansion of the pressure in powers of the density. Accordingly, we define the coefficients $c(\bsn)$ by the equation
\be
\label{def virial}
p(\bsz) = \sum_{\bsn} c(\bsn) \bsrho(\bsz)^{\bsn}.
\ee
Here also, we use the notation $\bsrho(\bsz)^{\bsn} = \prod_{i} \rho_{i}(\bsz)^{n_{i}}$. Let us check that the coefficients $c(\bsn)$ are well-defined in the sense of formal series. Observe that $[\bsz^{\bsk}] \bsrho(\bsz)^{\bsn} \neq 0$ only if $\bsn \leq \bsk$, i.e., if $n_{i} \leq k_{i}$ for all $i$; the notation $[\bsz^\bsk] \bsrho(\bsz)^\bsn$ refers to the coefficient of $\bsz^\bsk$ in the formal power series $\bsrho(\bsz)^\bsn$. Then $(\bsrho(\bsz)^{\bsn})_{\bsn}$ is a ``summable family'' of formal series in the sense of Def.~2.2 in \cite{ehrenborg-mendez94}, and the coefficients of $\bsz^{\bsk}$ in \eqref{def virial} satisfy
\be
\label{finite sum}
b(\bsk) = [\bsz^{\bsk}] \sum_{\bsn} c(\bsn) \bsrho(\bsz)^{\bsn} = \sum_{\bsn \leq \bsk} c(\bsn) [\bsz^{\bsk}] \bsrho(\bsz)^{\bsn},
\ee
the latter sum being finite. Then Eq.\ \eqref{finite sum} can be inverted recursively, and $c(k_{1}, \dots, k_{\ell}, 0, \dots)$ can be expressed in terms of $b(\bsn)$ and $c(n_{1}, \dots, n_{\ell}, 0, \dots)$ with $n_{i} \leq k_{i}$, $i=1,\dots, \ell-1$, and $n_{\ell} < k_{\ell}$.

Our goal is to control the convergence of the virial expansion, assuming convergence of the series $p(\bsz)$. In the following the statement ``$|\log f(\bsz)| \leq a$'' means that there is a $A(\bsz)\in \bbC$ such that $f(\bsz) = \exp(A(\bsz))$ and $|A(\bsz) |\leq a$, i.e., the precise choice of the branch of the logarithm is irrelevant. 

\begin{theorem}
\label{thm virial}
	Assume that there exist $0 < r_{i} < R_{i}$ and $a_{i}\geq0$, $i=1,2,\dots$, such that
	\begin{itemize}
		\item $p(\bsz)$ converges absolutely in the polydisk $D = \{\bsz \in \bbC^\bbN\mid \forall i \in\bbN:\, |z_{i}| < R_{i}\}$.
		\item $\displaystyle \Bigl| \log \frac{\partial p}{\partial z_{i}} (\bsz) \Bigr| < a_i$ for all $i\geq1$ and all $\bsz \in D$.
		\item $\displaystyle \sum_{i\geq1} \sqrt{\frac{r_{i}}{R_{i}}} < \infty$ and $\displaystyle \sum_{i\geq1} \frac{r_{i} a_{i}^{2}}{R_{i}} < \infty$.
	\end{itemize}
	Then there exists a constant $C<\infty$ (which depends on the $r_{i}$, $R_{i}$, $a_{i}$, but not on $\bsn$) such that
\be
|c(\bsn)| \leq C \sup_{\bsz \in D} |p(\bsz)| \prod_{i\geq1} \Bigl( \frac{\e{a_i}}{r_{i}} \Bigr)^{n_{i}}.
\ee
\end{theorem}

The estimate for $c(\bsn)$ guarantees convergence of the series $\sum_{\bsn} c(\bsn) \bsrho^{\bsn}$ for all $\bsrho$ in a polydisk 
\begin{equation} 
  D' = \Bigl\{\bsrho \in \bbC^\bbN \mid \forall i \in \bbN:\, |\rho_{i}| <  r_{i}\e{-a_i},\ 	\sum_{i \in \bbN} |\rho_i| \frac{\e{a_i}}{r_i} <\infty\Bigr \}. 
\end{equation} 	
Theorem \ref{thm virial} is proved in Section \ref{sec Lagrange-Good}.

We can also address the following question: Consider the functions $z_{i}(\bsrho)$ obtained by inverting \eqref{def densities}; for given $\bsrho \in D'$, does $\bsz(\bsrho)$ belong to $D$, so that $p(\bsz(\bsrho))$ is given by an absolutely convergent series? The following result provides a partial answer, as it guarantees convergence when $\bsrho$ belongs to a smaller domain.

\begin{theorem}
	\label{thm bound z}
	Under the same assumptions as in Theorem \ref{thm virial}, we have
	\[ 
		\Bigl| [\bsrho^{\bsn}] \frac{ \bsz(\bsrho)^{\bsk}}{\bsrho^{\bsk}} \Bigr| < C  \prod_{i\geq1} \frac{\e{a_i (n_{i}+k_{i})}}{r_i^{n_{i}}}.
	\]
\end{theorem}


The constant $C$ is the same as in Theorem \ref{thm virial}, and the proof is similar. It can be found at the end of Section \ref{sec Lagrange-Good}.

Let $i\in\bbN$ and choose $\bsk = (k_{j}) = (\delta_{i,j})$ in Theorem~\ref{thm bound z}.  We get that for all $\bsrho \in D'$, 
\be \label{eq:zrhobound}
	|z_{i}(\bsrho)| \leq C |\rho_i| \e{a_i} \prod_{j\geq1} \Bigl( 1 - \frac{\e{a_j} |\rho_{j}|}{r_{j}} \Bigr)^{-1},
\ee
so that $\bsz(\bsrho) \in D$ for $\bsrho$ small enough. The  inequality~\eqref{eq:zrhobound} is the analogue of the bound $|\rho_k(\bsz)|\leq |z_k| \exp(a_k)$, valid for $\bsz\in D$  under the assumptions of Theorem \ref{thm virial}.

\subsection{The point of view of the inverse function theorem} \label{sec:inverse-function}
The formulation of Theorem~\ref{thm virial} is geared towards the virial expansion in statistical mechanics. The theorem in itself is, however, purely analytic. In this section we rephrase it as a type of inverse function theorem and discuss its relation to traditional inverse function theorems.

	Let $\bigl(F_k(\boldsymbol{w})\bigr)_{k\in \bbN}$ be a family of power series in the complex variables $w_j$ ($j\in \bbN$) such that $F_k(0) \neq 0$, for all $k\in \bbN$.  The goal is to invert the system of equations $u_k = w_k F_k(\boldsymbol{w})$. On the level of formal power series, the inversion is always possible: there is a unique family of power series $G_k(\boldsymbol{u})$, $k \in \bbN$, 
such that the inverse is given by $w_k (\boldsymbol{u})= u_k G_k(\boldsymbol{u})$, i.e., for all $k \in \bbN$, we have  $w_k(\boldsymbol{u}) F_k( \boldsymbol{w}(\boldsymbol{u})) = u_k$,  as an identity of formal power series.

\begin{theorem} \label{thm:inverse-function}
	Assume that there exist $0 < r_{i} < R_{i}$ and $a_{i}\geq0$, $i=1,2,\dots$, such that
	\begin{itemize}
	\item The series $F_k(\boldsymbol{w})$, $k \in \bbN$, converge absolutely in the polydisk 	$D = \{\boldsymbol{w} \in \bbC^\bbN\mid \forall i \in\bbN:\, |w_{i}| < R_{i}\}$.
	\item $| \log F_i(\boldsymbol{w})| < a_i$ for all $i\geq1$ and all $\boldsymbol{w} \in D$.
\item $\displaystyle \sum_{i\geq1} \sqrt{\frac{r_{i}}{R_{i}}} < \infty$ and $\displaystyle \sum_{i\geq1} \frac{r_{i} a_{i}^{2}}{R_{i}} < \infty$.
\end{itemize}
Then there exists a constant $C<\infty$ (which depends on the $r_{i}$, $R_{i}$, $a_{i}$, but not on $\bsn$) such that for all $k \in \bbN$, 
\[
	\bigl| [\boldsymbol{u}^{\bsn}] G_k(\boldsymbol{u}) \bigr| \leq C \e{a_k} \prod_{i\geq1} \Bigl( 				\frac{\e{a_i}}{r_{i}} \Bigr)^{n_{i}}.
\]
\end{theorem} 
The proof is similar to the proofs of Theorems~\ref{thm virial} and~\ref{thm bound z} and it is omitted.
In order to better understand the analytic structure of the theorem, it is convenient to introduce Banach spaces of complex sequences. To simplify matters, suppose that the first two assumptions of the theorem hold with $a_k= ak $, where $a>0$ is some $k$-independent constant, and $R_k = \exp( - ak)$. Situations of this type occur in the context of cluster expansions. Choose $r_k:= k^{-4} \exp( - ak)$. 
Define the weighted $\ell^1$-norms
$\|\boldsymbol{w}\|_{p,q}:= \sum_{k\in \bbN} k^{p} \exp( q a k) |w_k|$. Let us define
\be
B_{p,q}(\eps):=\{ \boldsymbol{w} \in \bbC^\bbN:\, \|\boldsymbol{w}\|_{p,q} <\eps \}.
\ee 
Fix $\eps\in (0,1)$ small enough so that $C\eps/(1-\eps) <1$ and set  $c_\eps:= (1-\eps)^{-1} \exp(-\eps)$.  
As a consequence of Theorem \ref{thm:inverse-function} we get the well-defined functions
\begin{align}
\boldsymbol{f}:\ &B_{0,1}(1) \to B_{0,0}(1) &\quad \boldsymbol{g}:\ &B_{4,2}(\eps) \to B_{4,1}\bigl(C\eps/(1-\eps) \bigr)\subset B_{0,1}(1) \nonumber\\
&\boldsymbol{w} \mapsto \bigl( w_k F_k(\boldsymbol{w})\bigr)_{k\in \bbN} &&\boldsymbol{u} \mapsto \bigl( u_k G_k(\boldsymbol{u})\bigr)_{k\in \bbN}.
\end{align}
Then for every $\boldsymbol{u} \in B_{4,2}(\eps)$, we have
\be
\boldsymbol{f}\bigl(\boldsymbol{g}(\boldsymbol{u})\bigr)=\boldsymbol{u}.
\ee

It is instructive to compare this with traditional inverse function theorems: suppose that $\boldsymbol{f}$, considered as map from the Banach space with norm $\|\cdot\|_{4,1}$ to the space with norm $\|\cdot\|_{4,2}$, was Fr{\'e}chet-differentiable with invertible derivative in a neighbourhood of the origin. Then we would get the existence of open neighbourhoods of the origin such that $\boldsymbol{f}$ is a bijection between these neighbourhoods~\cite[Chapter 4.10]{zeidler}. 
Our result yields a weaker conclusion: we have a bijection between $B_{4,2}(\eps)$ and $\boldsymbol{g}(B_{4,2}(\eps))$, but in general the latter set needs not be open with respect to $\|\cdot\|_{4,1}$. 
The reason is that Theorem~\ref{thm:inverse-function} operates under conditions that are weaker than those of the inverse function theorem: for infinitely many variables, the existence of continuous partial derivatives $\partial \boldsymbol{f}/\partial w_k$ does not imply that $\boldsymbol{f}$ is Fr{\'e}chet-differentiable --- we do not know whether the Jacobi matrix $(\partial f_\ell/\partial w_k)$ represents a bounded operator. Our condition $  \exp(-a_k) \leq | f_k(\boldsymbol{w})/w_k| \leq \exp(a_k)$ replaces the traditional condition that the derivative and its inverse are bounded operators between Banach spaces.

\subsection{Lagrange-Good inversion \& bounds of virial coefficients}
\label{sec Lagrange-Good}

The Lagrange-Good inversion formula gives explicit expressions for $c(\bsn)$ and $[\bsrho^{\bsn}] \bsz^{\bsk}$, which can then be estimated. Let $N$ be the largest index $i$ such that $n_{i} \neq 0$, and consider the $N\times N$ matrix
\be
M(\bsz) =  \biggl( \delta_{ij} + \frac{z_{i} \frac{\partial^{2} p}{\partial z_{i} \partial z_{j}}}{\frac{\partial p}{\partial z_{i}}} \biggr)_{1\leq i,j \leq N}.
\ee
We use Eq.\ (4.5) of \cite{Ges} to get
\be
\label{Lagrange-Good}
[\bsrho^{\bsn}] \Phi(\bsz(\bsrho)) = [\bsz^{\bsn}] \Phi(\bsz) \frac1{\bigl( \frac{\partial p}{\partial\bsz} \bigr)^{\bsn}} \det M(\bsz).
\ee
Here, we used the notation $\bigl( \frac{\partial p}{\partial\bsz} \bigr)^{\bsn} = \prod_{i} \bigl( \frac{\partial p}{\partial z_{i}} \bigr)^{n_{i}}$.
We employ the formula above with $\Phi(\bsz) = p(\bsz)$ and $\Phi(\bsz) = \bsz^{\bsk} / \bsrho(\bsz)^{\bsk}$. 
In~\cite{Ges}, the formula has been proved for finitely many species; a proof for infinitely many species is given in~\cite{ehrenborg-mendez94} (Theorem 4). Note, however, that 
we can apply the finitely many species version in our context because we only need it for $\bsn$ with finitely many non-zero entries. 

In order to estimate the coefficients of the right side, we use Cauchy's formula and get upper bounds on the various terms. We start with the determinant in \eqref{Lagrange-Good}.

\begin{lemma}
\label{lem bound det}
Under the assumptions of Theorem \ref{thm virial}, there exists a constant $C<\infty$ such that for all $\bsz$ with $|z_{i}| = r_{i}$, and all $N$, we have
\[
|\det M(\bsz)| \leq C.
\]
\end{lemma}

\begin{proof}
We start by expanding $\det M(\bsz)$ in terms of determinants of minors. With $\caS_{N}$ and $\caS(J)$ denoting the set of permutations on $\{1,\dots,N\}$ and on $J \subset \{1,\dots,N\}$, we have
\be
\begin{split}
\det M(\bsz) &= \sum_{\sigma \in \caS_{N}} {\rm sgn}(\sigma) \prod_{i=1}^{N} \biggl( \delta_{i,\sigma(i)} + \frac{z_{i} \frac{\partial^{2} p}{\partial z_{i} \partial z_{\sigma(i)}}}{\frac{\partial p}{\partial z_{i}}} \biggr) \\
&= \sum_{\sigma \in \caS_{N}} {\rm sgn}(\sigma) \sum_{J \subset \{1,\dots,N\}} \prod_{i \in J^{\rm c}} \delta_{i,\sigma(i)} \prod_{i \in J} \frac{z_{i} \frac{\partial^{2} p}{\partial z_{i} \partial z_{\sigma(i)}}}{\frac{\partial p}{\partial z_{i}}} \\
&= \sum_{J \subset \{1,\dots,N\}} \sum_{\sigma \in \caS(J)} {\rm sgn}(\sigma) \prod_{i \in J} \frac{z_{i} \frac{\partial^{2} p}{\partial z_{i} \partial z_{\sigma(i)}}}{\frac{\partial p}{\partial z_{i}}} \\
&= \sum_{J \subset \{1,\dots,N\}} \det \biggl( \frac{z_{i} \frac{\partial^{2} p}{\partial z_{i} \partial z_{j}}}{\frac{\partial p}{\partial z_{i}}} \biggr)_{ i,j \in J},
\end{split}
\ee
where the summand corresponding to $J= \emptyset$ is by definition equal to $1$.
Let $u_{i}$ be non-zero numbers to be determined later. We use the identity $\det M = \det D M D^{-1}$ with $D$ the diagonal matrix with entries $u_{i}^{-1} \frac{\partial p}{\partial z_i}$ in the diagonal, and we get 
\be
\det M(\bsz) = \sum_{J \subset \{1,\dots,N\}} \det \biggl( z_{i} \frac{u_{j}}{u_{i}} \frac{\partial}{\partial z_{i}} \log \frac{\partial p}{\partial z_{j}} \biggr)_{ i,j \in J}.
\ee
From Hadamard's inequality, we get the upper bound
\be
|\det M(\bsz)| \leq \sum_{J \subset \{1,\dots,N\}} \prod_{i\in J} \biggl( \sum_{j\in J} |z_{i}|^{2} \frac{u_{j}^{2}}{u_{i}^{2}} \Bigl| \frac{\partial}{\partial z_{i}} \log \frac{\partial p}{\partial z_{j}} \Bigr|^{2} \biggr)^{1/2}.
\ee
By Cauchy's formula, choosing the contour around $z_{i}$ with radius $R_{i} - r_{i}$, and using the bound on the logarithm of $\frac{\partial p}{\partial z_{j}}$, we get
\be
\Bigl| \frac{\partial}{\partial z_{i}} \log \frac{\partial p}{\partial z_{j}}(\bsz) \Bigr| = \biggl| \frac1{2\pi\ii} \oint \frac{\log \frac{\partial p}{\partial z_{j}}(\hat\bsz^{(i)},w)}{(w-z_{i})^{2}} \dd w \biggr| \leq \frac{a_j}{R_{i} - r_{i}},
\ee
where $\hat\bsz^{(i)}$ is the vector $\bsz$ without the $z_{i}$ term. Then
\be
\begin{split}
|\det M(\bsz)| &\leq \sum_{J \subset \{1,\dots,N\}} \prod_{i\in J} \biggl[ \frac{r_{i}}{u_{i} (R_{i}-r_{i})} \Bigl( \sum_{j\in J} u_{j}^{2} a_{j}^{2} \Bigl)^{1/2} \biggr] \\
&\leq \prod_{i=1}^{N} \biggl[ 1 + \frac{r_{i}}{u_{i} (R_{i}-r_{i})} \Bigl( \sum_{j\geq1} u_{j}^{2} a_{j}^{2} \Bigl)^{1/2} \biggr] \\
&\leq \exp \biggl[ \sum_{i\geq1} \frac{r_{i}}{u_{i} (R_{i}-r_{i})} \Bigl( \sum_{j\geq1} u_{j}^{2} a_{j}^{2} \Bigl)^{1/2} \biggr].
\end{split}
\ee
Choosing $u_{j} = \sqrt{r_{j} / R_{j}}$, the expression above is finite by the assumptions of the lemma.
\end{proof}

\begin{proof}[Proof of Theorem \ref{thm virial}]
We use the Lagrange-Good inversion \eqref{Lagrange-Good} and Cauchy's formula and we write
\be
c(\bsn) = \biggl( \prod_{i=1}^{N} \frac1{2\pi\ii} \oint \frac{\dd z_{i}}{z_{i}^{n_{i}+1}} \biggr) p(\bsz) \frac1{\bigl( \frac{\partial p}{\partial\bsz}(\bsz) \bigr)^{\bsn}} \det M(\bsz),
\ee
with the contours being circles of radii $r_{i}$ around the origin. It follows from the assumptions that $\bigl| \frac{\partial p}{\partial z_{i}} \bigr| > \e{-a_i}$ for all $i$ and all $\bsz \in D$. Theorem \ref{thm virial} then follows from Lemma \ref{lem bound det}.
\end{proof}

\begin{proof}[Proof of Theorem \ref{thm bound z}]
We use Eq.\ \eqref{Lagrange-Good} with $\Phi(\bsz) = \bsz^{\bsk}/\bsrho(\bsz)^\bsk$. 
We then repeat the proof of Theorem \ref{thm virial} without the term $p(\bsz)$. This yields the bound
\be
\begin{split} 
	\Bigl|[\bsrho^{\bsn}] \frac{\bsz(\bsrho)^{\bsk}}{\bsrho^{\bsk}} \Bigr|
	&= \Bigl|[\bsz^{\bsn}] \frac{\bsz^{\bsk}}{\bsrho(\bsz)^{\bsk}} 
	\frac{1}{\bigl( \frac{\partial p}{\partial \bsz}\bigr)^{\bsn}}  \det M(\bsz) \Bigr| \\
	&= \Bigl|[\bsz^{\bsn}] \frac{1}{\bigl( \frac{\partial p}{\partial \bsz}\bigr)^{\bsn+\bsk}}  \det M(\bsz) \Bigr| \\
	& < C  \prod_{i \geq 1} \frac{\e{a_i (n_{i}+k_{i})}}{r_i^{n_{i}}}.
\end{split}
\ee
\end{proof}

\section{Connected and two-connected graphs}
\label{sec graphs}

We now make a further assumption on the series $p(\bsz)$ introduced in \eqref{p(z)}: it is given by the (weighted) exponential generating function of coloured graphs. This choice is motivated by applications to statistical mechanics, which are discussed in Section \ref{sec clexp}. If the weights satisfy a certain {\it block factorisation} property, the virial coefficients $c(\bsn)$ can be expressed using two-connected graphs. We therefore introduce these coloured graphs in some detail.

Recall that a graph $g$ is a pair $g=(\mathcal{V}(g),\mathcal{E}(g))$ with $\mathcal{V}(g)$ a set and 
$\mathcal{E}(g) \subset \bigl\{ \{i,j\} \subset \caV(g) : i \neq j \bigr\}$. The elements of $\mathcal{V}(g)$ are called ``vertices'' and the elements of $\mathcal{E}(g)$ ``edges''.
A graph $g'$ is a \emph{subgraph} of $g$ if $\mathcal{V}(g') \subset \mathcal{V}(g)$ and $\mathcal{E}(g') \subset\mathcal{E}(g)$, in which case we write $g'\subset g$. 
A graph $g$ is {\it connected} if $|\mathcal{V}(g)|\geq 1$ and for every partition $\caV(g) = \caV_{1} \cup \caV_{2}$, there exists $i \in \caV_{1}$ and $j \in \caV_{2}$ such that $\{i,j\} \in \mathcal{E}(g)$. A graph $g$ is {\it two-connected} if $|\mathcal{V}(g)|\geq 2$ and for every $i \in \caV(g)$, the subgraph $g \setminus \{i\}$, obtained by removing $i$ and all edges containing $i$, is connected. 

A coloured graph  is a pair $(g,\bsk)$ where $g$ is a graph and  and  $\bsk$ is a function $\caV(g) \to \bbN$ that assigns the colour $k_{i} \in \bbN$ to each $i \in \caV(g)$. Coloured connected graphs are pairs $(g,\bsk)$ with $g$ a connected graph, coloured two-connected graphs are pairs $(g,\bsk)$ with $g$ a two-connected graph. The results of this section can be formulated in the general framework of \emph{labelled coloured combinatorial species} (\cite{mendez-nava93}, \cite[Sections 2 and 3]{ehrenborg-mendez94}), but for the reader's convenience we present the results in a self-contained way.
 
Let $w(g,\bsk)$ be a weight function on coloured graphs. We assume that it is invariant under relabellings that preserve the colour: let $\sigma:\mathcal{V}\to \mathcal{V}$ be a bijection with the property that $k_{\sigma(i)} = k_{i}$ for all $i\in \mathcal{V}$, and $g_\sigma$ the graph with vertices $\mathcal{V}(g_\sigma)= \mathcal{V}(g)$ and edges $\mathcal{E}(g_\sigma) = \bigl\{ \{\sigma(i),\sigma(j)\}:\ \{i,j\} \in \mathcal{E}(g)\bigr \}$, then $w(g_\sigma,\bsk) =w(g,\bsk).$  The weighted exponential generating function for connected graphs is defined by
\be
\label{def gen fct conn}
\begin{split}
C(\bsz) &= \sum_{n\geq1} \frac1{n!} \sum_{\bsk = (k_{1}, \dots, k_{n})} z_{k_{1}} \dots z_{k_{n}} \sum_{g \in \caC_{n}} w(g,\bsk) \\
&= \sum_{\bsn} \frac{\bsz^{\bsn}}{\bsn!} \sum_{\bsg \in \caC[\bsn]} w(\bsg).
\end{split}
\ee
In the first line we denoted by $\caC_{n}$ the set of connected graphs with vertex set $\{1,\ldots,n\}$. In the second line the sum is over multi-indices $\bsn$ with finitely many entries, and $\caC[\bsn]$ denotes the set of coloured graphs $\bsg=(g,\bsk^{[\bsn]})$ with vertices $1,2,\dots, |\bsn|$ 
and $\bsk^{[\bsn]}$ the colouring 
 such that the first $n_{1}$ vertices have colour 1, the vertices $n_{1}+1, \dots, n_{1} + n_{2}$ have colour 2, etc. We shall refer to $\bsk^{[\bsn]}$ as the \emph{canonical colouring}. 
The second expression for $C(\bsz)$ is more elegant but the first expression turns out to be more practical. These formul\ae{} should still be understood as formal series.

We suppose that $w(g,\bsk)$ factorises with respect to the block decomposition of $g$. 
Recall that an {\it articulation point} of $g$ is a vertex $i \in \caV$ such that the subgraph $g \setminus \{i\}$ is disconnected. (A two-connected graph is a connected graph without articulation point.)
A \emph{block} is a two-connected subgraph $g'$ of $g$ that is maximal, i.e., if $g''$ is   a two-connected subgraph  such that $g'\subset g'' \subset g$, then $g''=g'$. 
Let $\{g_1,\ldots,g_m\}$ be the \emph{block decomposition} of $g$, i.e., the set of blocks of $g$. It has the following properties.
(i) The blocks induce a partition of the edge sets:
$\mathcal{E}(g)= \cup_{i=1}^m \mathcal{E}(g_i)$ with  $\mathcal{E}(g_i)\cap \mathcal{E}(g_j) =\emptyset$ when $i\neq j$. (ii) Each articulation point belongs to more than one $\caV(g_i)$, other vertices belong to exactly one $\caV(g_{i})$. 
\begin{figure}[here]
	\begin{minipage}[b]{0.75\linewidth}
		\centering
		\includegraphics[scale=0.27]{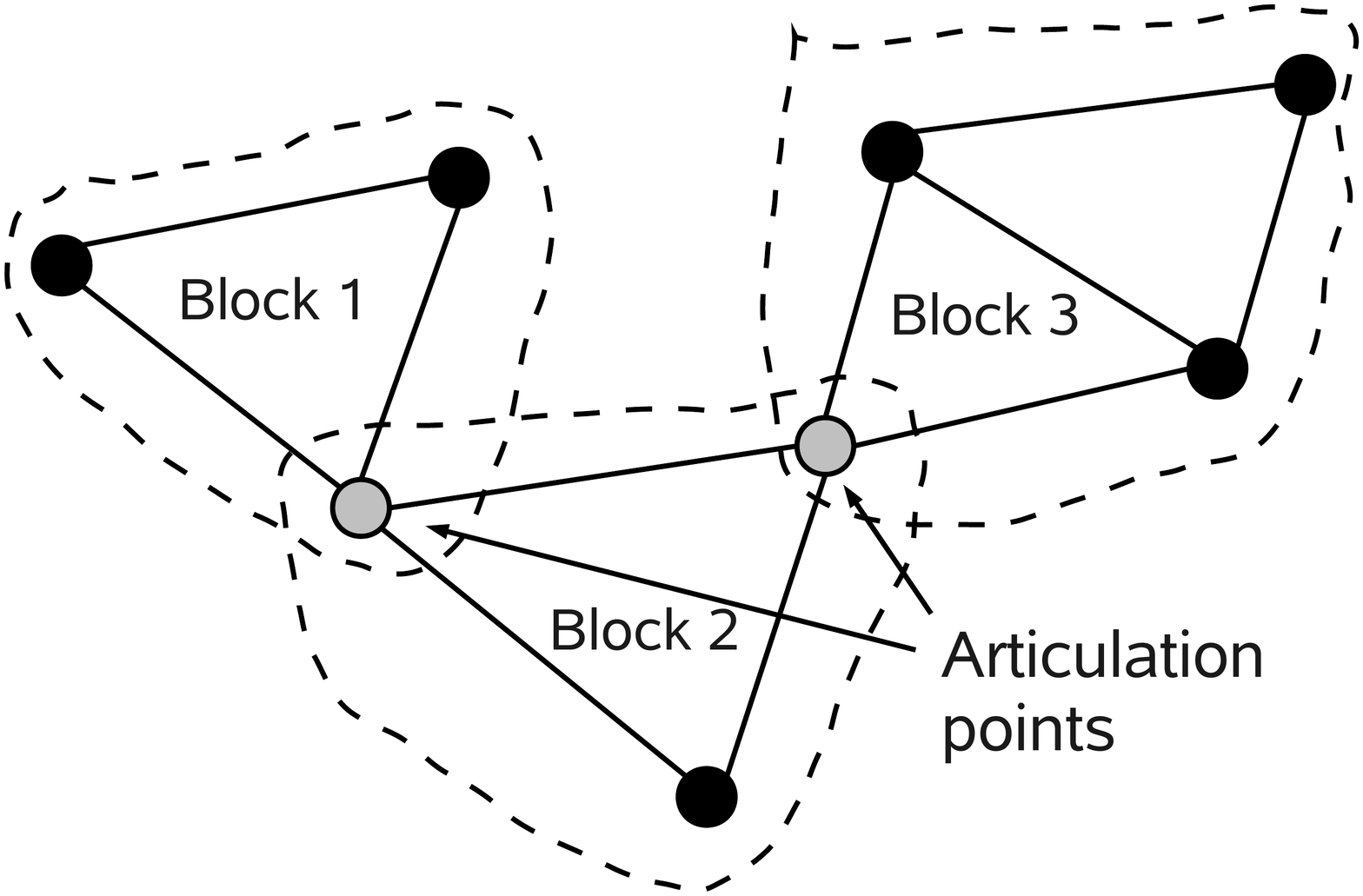}
		\caption{The block decomposition of a connected graph.}
		\label{fig:blocks}
	\end{minipage}
\end{figure}

\begin{theorem}
\label{thm 2-virial}
Assume that $p(\bsz) = C(\bsz)$ as above, that the weight function $w(g,\bsk)$ satisfies 
$w(g,\bsk)=1$ when $g$ has size $n=1$, and 
\[
w(g,\bsk) = \prod_{i=1}^{m} w(g_{i},\bsk\bigl|_{\mathcal{V}(g_i)})
\]
when $g$ has size $n\geq 2$, 
where $\{ g_{1}, \dots, g_{m} \}$ is the block decomposition of $g$ and $\bsk\bigl|_{\mathcal{V}(g_i)}$ is the restriction of the colouring $\bsk:\mathcal{V}(g)\to \bbN$ to  $\mathcal{V}(g_i)$.
 Then
\[
p(\bsz) = \sum_{k\geq1} \rho_{k}(\bsz) - \sum_{\bsn:\, |\bsn|\geq 2} \bigl( |\bsn| - 1 \bigr) \frac{\bsrho(\bsz)^{\bsn}}{\bsn!} \sum_{\bsg \in \caB[\bsn]} w(\bsg)
\]
where $\mathcal{B}[\bsn]$ consists of the two-connected coloured graphs $\bsg = (g,\bsk^{[\bsn]})$ with canonical colouring $\bsk^{[\bsn]}$ and vertex set $\mathcal{V}(g) = \{1,2,\ldots,|\bsn|\}$. 
\end{theorem}

The proof is given at the end of this section. It uses the dissymmetry theorem of combinatorial structures, following \cite{BLL,L}. This is rather straightforward but we need to add colours to all objects.

If $k \in \bbN$ denotes a colour, a $k$-rooted graph is a triplet $(g,\bsk,i)$ where $g$ is a graph with finite vertex set $\mathcal{V}(g)$, $\bsk \in \bbN^{\caV(g)}$, 
$i \in \caV(g)$, with the property that the root $i$ has colour $k_{i}=k$. The weighted exponential generating function of $k$-rooted connected graphs is
\be
\begin{split}
C^{\bullet k}(\bsz) &= \sum_{n\geq1} \frac1{n!} \sum_{\bsk=(k_{1},\dots,k_{n})} z_{k_{1}} \dots z_{k_{n}} \bigl| \{ i : k_{i} = k \} \bigr| \sum_{g \in \caC_{n}} w(g,\bsk) \\
&= \sum_{\bsn} n_{k} \frac{\bsz^{\bsn}}{\bsn!} \sum_{\bsg \in \caC[\bsn]} w(\bsg) \\
&= z_k \frac{\partial C}{\partial z_k} (\bsz).
\end{split} 
\ee

Let $\mathcal{C}$, $\mathcal{C}^{\bullet k}$ be the sets of connected (resp. $k$-rooted connected) coloured graphs with vertex set of the form $\{1,\ldots,n\}$, $n\in \bbN$, and set and $\mathcal{C}^\bullet:=\cup_{k\in \bbN} \mathcal{C}^{\bullet k}$. The associated exponential generating function is $C^{\bullet}(\bsz) = \sum_{k \in \bbN} C^{\bullet k}(\bsz)$. We define $\mathcal{B}$, $\mathcal{B}^{\bullet}$ and their exponential generating functions $B(\bsz)$, $B^\bullet(\bsz)$ in a similar way, replacing ``connected'' by ``two-connected''. 

 Next, we describe the composition of connected and two-connected graphs. 
The set $\caB( \{ \caC^{\bullet k} \}_{k\in\bbN})$ consists of coloured two-connected graphs whose vertices contain a rooted connected graph with the appropriate colour for the root. More precisely, an element $\bsg \in \caB( \{ \caC^{\bullet k} \}_{k\in\bbN})$ of size $n$ is a triple $\boldsymbol{\gamma} = (\bsk,\gamma,(\gamma_i)_{i\in \mathcal{V}(\gamma)})$ consisting of
\begin{itemize}
	\item A colour assignment $\bsk \in \bbN^{n}$.
	\item A two-connected graph $\gamma$ with vertex set $\mathcal{V}(\gamma) \subset \{1,\ldots,n\}$, $|\mathcal{V}(\gamma)|\geq 2$. 
	\item A family $(\gamma_i)_{i\in \mathcal{V}(\gamma)}$ of connected graphs $\gamma_i$ such that 
			$i \in \mathcal{V}(\gamma_i)$, and the vertex sets form a partition $\{1,\ldots,n\} = \cup_{i\in \mathcal{V}(\gamma)} \mathcal{V}(\gamma_i)$. Note that $(\gamma_i, \bsk|_{\mathcal{V}(\gamma_i)}, i)$ is a $k_i$-rooted coloured connected graph.
\end{itemize}
 With each $\boldsymbol{\gamma}$ we associate the connected graph $g=g(\boldsymbol{\gamma})$ with vertices $1,\ldots,n$ and edge set $\mathcal{E}(\gamma) \cup \bigl(\cup_{i\in \mathcal{V}(\gamma)} \mathcal{E}(\gamma_i)\bigr)$. We assign to the composite structure $\boldsymbol{\gamma}$ the weight $w(g(\boldsymbol{\gamma}),\bsk)$ of the underlying connected coloured graph. 
We also introduce $\caB^{\bullet}( \{\caC^{\bullet k} \}_{k\in\bbN})$, which is as above but with the additional choice of a root in $\caV(\gamma)$.

\begin{lemma}
\label{lem compositions} Under the assumptions of Theorem~\ref{thm 2-virial}, 
the weighted exponential generating functions of $\caB( \{ \caC^{\bullet k} \}_{k\in\bbN})$ and $\caB^{\bullet}( \{\caC^{\bullet k} \}_{k\in\bbN})$ satisfy
\[
\begin{split}
&B( \{ C^{\bullet k}(\bsz) \}_{k\in\bbN}) = \sum_{\boldsymbol{\gamma} \in \caB( \{ \caC^{\bullet k} \}_{k\in\bbN})} \frac1{n!} z_{k_{1}} \dots z_{k_{n}} w(g,\bsk), \\
&B^{\bullet}( \{ C^{\bullet k}(\bsz) \}_{k\in\bbN}) = \sum_{\boldsymbol{\gamma} \in \caB^{\bullet}( \{ \caC^{\bullet k} \}_{k\in\bbN})} \frac1{n!} z_{k_{1}} \dots z_{k_{n}} w(g,\bsk).
\end{split}
\]
Here, $n = n(\bsg)$ is the size of $\bsg$, $\bsk = \bsk(\bsg)$ is the colour assignment of the vertices $\{1,\dots,n\}$ and $g=g(\boldsymbol{\gamma})$ is the induced connected graph on the set of vertices $\{1,\dots,n\}$.
\end{lemma}

The proofs are tedious but immediate: one sums over all components of $\bsg$, and uses the multinomial theorem so that the elements of the partition become independent. This is possible because of the factorisation property of the weights,
\begin{equation} 
	w(g(\boldsymbol{\gamma}),\bsk) = w(\gamma, \bsk|_{\mathcal{V}(\gamma)}) \times \prod_{i\in \mathcal{V}(\gamma)} w(\gamma_i,\bsk|_{\mathcal{V}(\gamma_i)}).
\end{equation}
For a proof in the context of labelled coloured combinatorial species (but without weights), see~\cite[Proposition 1.4]{mendez-nava93} and~\cite[Proposition 1.3]{ehrenborg-mendez94}.

Next, we state the dissymmetry theorem for coloured graphs. Here, $A+B$ denotes the {\it disjoint union} where the elements of $A$ and $B$ are distinct by definition.

\begin{theorem} We have
\[
\caC + \caB^{\bullet}( \{ \caC^{\bullet k} \}_{k\in\bbN} ) = \caC^{\bullet} + \caB( \{ \caC^{\bullet k} \}_{k\in\bbN} )
\]
in the sense that there is a size and weight preserving bijection between $\caC + \caB^{\bullet}( \{ \caC^{\bullet k} \}_{k\in\bbN} )$ and $\caC^{\bullet} + \caB( \{ \caC^{\bullet k} \}_{k\in\bbN} )$.
\end{theorem}

\begin{proof}
There are two mappings $\phi: \caC + \caB^{\bullet}( \{ \caC^{\bullet k} \}_{k\in\bbN} ) \to \mathcal{C}$ and $\psi:\caC^{\bullet} + \caB( \{ \caC^{\bullet k} \}_{k\in\bbN} )\to \mathcal{C}$, which associate to each graph in each of the sets above a unique connected graph. The idea is to informally `forget' the extra structure afforded to us by the composite structures. 

These two mappings are conveniently described in terms of their preimages (the structures corresponding to the same connected graph):

 The preimage of $g$ under $\phi$ consists of the union of: 
\begin{itemize}
	\item The set containing the graph itself, $g\in \mathcal{C}$.
	\item The set of composite structures $(\bsk,\gamma,(\gamma_i)_{i\in \mathcal{V}(\gamma)},r) \in 
			 \caB^{\bullet}( \{ \caC^{\bullet k} \}_{k\in\bbN} )$ where $\gamma$ is one of the blocks $g_1,\ldots,g_m$ of $g$, $r\in \mathcal{V}(\gamma)$, and $(\gamma_i)$ are uniquely determined by $g$ and the choice of $\gamma$. 
\end{itemize}
 
The preimage of $g$ under $\psi$ consists of the union of:

\begin{itemize}
	\item The set of ordered pairs $(g, i)_{i \in \{1, \cdots n\}} \in \caC^{\bullet}$, where the second entry indicates a root.
	\item The set of composite structures $(\bsk, \gamma, (\gamma_i)_{i \in \mathcal{V}(\gamma)})$, where $\gamma$ is one of the blocks $g_1,\ldots,g_m$ of $g$ and $(\gamma_i)$ are uniquely determined by $g$ and the choice of $\gamma$, as described above.
\end{itemize}

It is sufficient to prove that for every $\bsg=(g,\bsk) \in \mathcal{C}$, the preimages $\phi^{-1}(\{\bsg\})$ and $\psi^{-1} (\{\bsg\})$ have the same cardinality.

Let $\bsg=(g,\bsk) \in \mathcal{C}$. If $g$ has size one, then $\phi^{-1}(\{\boldsymbol{g}\}) = \{\boldsymbol{g}\}\subset \mathcal{C}$ and $\psi^{-1} (\{\boldsymbol{g}\}) = \{(g,\bsk,1)\} \subset \caC^{\bullet}$, and both preimages have cardinality $1$. 
If $g$ has size $n\geq 2$, let $\{g_{1},\dots,g_{m}\}$ be the block decomposition of $g$. 
 
The preimage under $\phi$ has cardinality 
\begin{equation} 
	1 +\sum_{i=1}^m |\mathcal{V}(g_i)|. 
\end{equation} 

The sum gives the possible roots for each block considered in turn.

The preimage under $\psi$ has cardinality 
\be
	n+m.
\ee
The first ``$n$'' corresponds to the number of ways to choose the root when the preimage is in $\caC^{\bullet}$, and $m$ is the number of composite structures $\boldsymbol{\gamma} \in 
\caB^{\bullet}( \{ \caC^{\bullet k} \}_{k\in\bbN} )$ with $g(\boldsymbol{\gamma}) = g$. 

There remains to show that the block decomposition of every $g \in \caC_{n}$ satisfies
\be
	1 + \sum_{i=1}^{m} |\caV(g_i)| = n + m,
\ee
or equivalently
\be
	\sum_{i=1}^{m} \bigl( |\caV(g_i)|-1 \bigr) = n-1.
\ee
This can be seen by induction. This clearly holds when $m=1$ and $\caV(g_1) = \{1,\dots,n\}$ (this corresponds to $g$ being two-connected). Now suppose that $g \in \caC_{n}$ has $m$ blocks of size $n_{1},\dots,n_{m}$. 
Consider the bipartite graph $t$ whose vertex set consists of the blocks $g_1,\ldots,g_m$ and articulation points $a_1,\ldots,a_\ell$ of $g$, and edges $\{ \{g_i, a_j\}: \ a_j \in g_i\}$. The graph $t$ is known~\cite[Section 4.3]{BLL} to be a tree and called the \emph{block cut-point tree} of $g$. Let $v \in \mathcal{V}(t)$ be a leaf of $t$. Then $v$ is  a vertex belonging to exactly one edge and must be a block $v=g_i$ containing exactly one articulation point $a$ of $g$. 

Thus there is a block containing precisely one articulation point of $g$, 
 and without loss of generality we take this to be the $m$th block. We remove from $g$ 
 all edges of the block $g_m$ and all vertices of $g_m$, except the articulation point $a$.   We now have a graph with $m-1$ blocks and so we have $\sum_{i=1}^{m-1} (n_{i}-1) = n - n_{m}$ by induction. Therefore,
\be
\sum_{i=1}^{m} (n_{i}-1) = n - n_{m} + n_{m} - 1 = n-1.
\ee
\end{proof}

\begin{proof}[Proof of Theorem \ref{thm 2-virial}]
Lemma \ref{lem compositions} and the dissymmetry theorem imply that the exponential generating functions satisfy
\be
C(\bsz) + B^{\bullet}( \{ C^{\bullet k}(\bsz) \}_{k\in\bbN} ) = C^{\bullet}(\bsz) + B( \{ C^{\bullet k}(\bsz) \}_{k\in\bbN} ).
\ee
To directly compare with Theorem~\ref{thm 2-virial}, we write this as:
\be
C(\bsz)= C^{\bullet}(\bsz) + B( \{ C^{\bullet k}(\bsz) \}_{k\in\bbN} )-B^{\bullet}( \{ C^{\bullet k}(\bsz) \}_{k\in\bbN} ).
\ee
We have $p(\bsz) = C(\bsz)$ and $\rho_{k}(\bsz) = C^{\bullet k}(\bsz)$, and the theorem follows.
\end{proof}

We conclude this section with two remarks. The first remark is that under the assumptions of Theorem~\ref{thm 2-virial}, we also have a formula for the expansion of the chemical potential $\log z_k$ in terms of the density, 
\be
	\log z_k = \log \rho_k(\bsz) - \frac{\partial B}{\partial \rho_k} \bigl( \bsrho(\bsz)\bigr),\quad B(\bsrho)= \sum_{\bsn:\,|\bsn|\geq 2}\frac{\bsrho^{\bsn}}{\bsn!}  \sum_{\bsg \in \mathcal{B}[\bsn]} w(\bsg).
\ee
This follows from the relation 
\be  \label{eq:bctree}
	\rho_k(\bsz) = C^{\bullet k}(\bsz) = z_k \exp\Bigl( \frac{\partial B}{\partial \rho_k} \bigl( \bsrho(\bsz)\bigr)\Bigr)
\ee
\cite[Section 3.2]{Far10b}; $\partial_k B$ is the generating function for two-connected graphs whose root is a ``ghost'' of colour $k$.

The second remark is that Theorem~\ref{thm 2-virial} is not limited to connected and biconnected graphs, but holds for pairs of combinatorial structures with a similar composition structure -- as Leroux puts it, for ``various tree-like structures''~\cite{L}. A well-known example is the dissymmetry theorem for trees~\cite[Chapter 4]{BLL}, which can be adapted to coloured trees with colour-dependent weights and constraints. This is interesting because Good's original motivation for his multi-variable version of the Lagrange inversion came from branching processes in probability and combinatorics of trees~\cite{good60,good65}.

\section{Classical gas of rigid molecules}
\label{sec clexp}

We now describe a physical system that fits the theory of Sections \ref{sec virial} and \ref{sec graphs}. It consists of a gas of molecules that are assumed to be rigid. Let $\Lambda \subset \bbR^{d}$ be the domain, which we take as a cube in $\bbR^{d}$ with periodic boundary conditions. We let $V$ denote its volume. A molecule is represented by
\be
X = (k,x,\phi),
\ee
where $k \in \bbN$ denotes the species, $x \in \Lambda$ denotes the position, and $\phi \in \Phi:= S^{d-1}=\{ \phi \in \bbR^d\mid \|\phi\|=1\}$ denotes the orientation. Interactions are given by a function $U(X_{1},X_{2})$ that takes values in $\bbR \cup \{+\infty\}$. Let
\be
\zeta(X_{1},X_{2}) = \e{-U(X_{1},X_{2})} - 1.
\ee
We take periodic boundary conditions, i.e., we assume that if $\Lambda=[0,L]^d$, and $y-y' \in L \mathbb{Z}^d$, then 
$U((x,k,\phi),(y,\ell,\psi)) = U((x,k,\phi),(y',\ell,\psi))$.
We make three assumptions on the interactions. The first one is about symmetries, the second one is the stability condition that ensures the existence of the thermodynamic limit, and the last one implies that we consider a regime of low density or high temperatures.

\begin{assumption}
\label{ass symmetries}
The potential function $U$ satisfies
\begin{itemize}
\item Symmetry: $U(X_{1},X_{2}) = U(X_{2},X_{1})$.
\item Translation invariance: If $X+a$ denotes the molecule translated by $a \in \Lambda$, i.e., with position $x+a$, then $U(X_{1}+a, X_{2}+a) = U(X_{1},X_{2})$.
\item Rotation invariance: If $RX$ denotes the molecule rotated by the orthogonal matrix $R$, i.e., with orientation $R\phi$, then $U(RX_{1}, R X_{2}) = U(X_{1},X_{2})$.
\end{itemize}
\end{assumption}

The {\it partition function} of the system is
\be
\label{def part fct}
Z_{\Lambda}(\bsz) = \sum_{n\geq0} \frac1{n!} \sum_{\bsk \in \bbN^{n}} z_{k_{1}} \dots z_{k_{n}} \int_{\Lambda^{n}} \dd x_{1} \dots \dd x_{n} \int_{\Phi^{n}} \dd\phi_{1} \dots \dd\phi_{n}
\exp \Bigl\{ -\sum_{1\leq i<j\leq n} U(X_{i},X_{j}) \Bigr\}.
\ee
The term $n=0$ is understood to be equal to 1, and $\dd \phi$ is the unique rotationally invariant measure on $\Phi$ with $\int_\Phi \dd \phi =1$.
 Given a graph $g \in \caC_{n}$, we define the weight function $w_{\Lambda}(g,\bsk)$ to be
\be
\label{def weight}
w_{\Lambda}(g,\bsk) = \frac1V \int_{\Lambda^{n}} \dd x_{1} \dots \dd x_{n} \int_{\Phi^{n}} \dd\phi_{1} \dots \dd\phi_{n} \prod_{\{i,j\} \in \mathcal{E}(g)} \zeta(X_{i},X_{j}),
\ee
where
\be
\zeta(X_{i},X_{j}) = \e{-U(X_{i},X_{j})} - 1.
\ee

The empty product is set to be equal to $1$, so that graphs of size $1$ have weight $V^{-1} \int_\Lambda \dd x \int_\Phi \dd \phi =1$.
By a standard cluster expansion, or by the exponential formula of combinatorial structures, we have
\be
Z_{\Lambda}(\bsz) = \exp \biggl\{ V \sum_{n\geq1} \frac1{n!} \sum_{\bsk \in \bbN^{n}} z_{k_{1}} \dots z_{k_{n}} \sum_{g\in\caC_{n}} w_{\Lambda}(g,\bsk) \biggr\}.
\ee
The partition function is related to the finite-volume pressure by 
$Z_{\Lambda} = \e{Vp_{\Lambda}}$. We then define
\be
p_{\Lambda}(\bsz) = C(\bsz),
\ee
where $C(\bsz)$ is the exponential generating function of connected graphs given in \eqref{def gen fct conn} with the weights $w_{\Lambda}(g,\bsk)$ in \eqref{def weight}. The goal is to show that the assumptions of Theorem \ref{thm virial} hold true uniformly in the volume $V$.

\begin{lemma}
\label{lem block factorisation}
Under Assumption \ref{ass symmetries}, the weight function of \eqref{def weight} satisfies the block factorisation
\[
w_{\Lambda}(g,\bsk) = \prod_{i=1}^{m} w_{\Lambda}(g_{i},\bsk).
\]
\end{lemma}

It is not too hard to check that factorisation holds when the graph is cut at any articulation point, and the lemma follows. It should be stressed that Lemma \ref{lem block factorisation} fails when the molecules are not assumed to be rigid. Next, the stability condition.

\begin{assumption}
\label{ass stability}
There exists a nonnegative constant $b$ such that for all $n$ and all $X_{1}, \dots, X_{n}$, we have
\be
\prod_{1\leq i<j\leq n} \bigl| 1 + \zeta(X_{i},X_{j}) \bigr| \leq \prod_{i=1}^{n} \e{b k_{i}}.
\ee
In addition, we also assume that for all $X$ and $Y$ of species $k$ and $\ell$, we have
\be
\label{strong KP}
\bigl| 1 + \zeta(X,Y) \bigr| \leq \e{b \min(k,\ell)}.
\ee
\end{assumption}

The next and last assumption is the ``Koteck\'y-Preiss criterion'' that guarantees that the interactions and the weights are small.

\begin{assumption}
\label{ass KP}
There exist positive numbers $R_{1}, R_{2}, \dots$ and a constant $a$ such that for all $X = (k,x,\phi)$,
\be
\sum_{k'\in\bbN} R_{k'} \e{(a+3b) k'} \int_{\bbR^{d}} \dd x' \int_{\Phi} \dd\phi' |\zeta(X,X')| \leq ak,
\ee
where $b$ is given in Assumption \ref{ass stability}.
In addition, we also assume that
\be
\sum_{k' \in \bbN} R_{k'} \e{(a+3b) k'} < \infty.
\ee
\end{assumption}

\begin{theorem}
Let $p_{\Lambda}(\bsz) = C(\bsz)$ and suppose that Assumptions \ref{ass symmetries}--\ref{ass KP} hold true. Then
\begin{itemize}
\item[(a)] $p_{\Lambda}(\bsz)$ converges absolutely in the polydisc $D = \{ \bsz\in\bbC^{\bbN} : |z_{i}| < R_{i} \; \forall i \in \bbN \}$.
\item[(b)] $\displaystyle \Bigl| \log \frac{\partial p_{\Lambda}}{\partial z_{k}} (\bsz) \Bigr| < a k$ for all $\bsz \in D$ and all $k \in \bbN$.
\end{itemize}
\end{theorem}

The main consequence of this theorem is that Theorem \ref{thm virial} applies, hence the existence of a domain of densities with absolute convergence of the virial expansion.

\begin{proof}
The setting of \cite{PU} applies directly here. The measure space of ``polymers'' $(\bbX,\mu)$ in \cite{PU} is presently given by $\bbX = \bbN \times \Lambda \times \Phi$ with $\mu$ the measure such that
\be
\int_{\bbX} f(X) \dd\mu(X) = \sum_{k \in \bbN} z_{k} \int_{\Lambda} \dd x \int_{\Phi} \dd\phi f(k,x,\phi)
\ee
for arbitrary integrable function $f$ on $\bbX$.

The conditions of \cite{PU} are fulfilled --- our Assumption \ref{ass KP} being slightly stronger with $3b$ instead of $2b$, but it will be needed in the proof of (b). From Theorem 2.1 in \cite{PU} we have that for every $X_{1} = (k_{1},x_{1},\phi_{1})$, and every $\bsz \in D$,
\be
\label{THE bound}
\begin{split}
\sum_{n\geq2} \frac1{(n-1)!} &\sum_{k_{2},\dots,k_{n} \in \bbN} |z_{k_{2}}| \dots |z_{k_{n}}| \int_{\bbR^{d}} \dd x_{2} \dots \int_{\bbR^{d}} \dd x_{n} \\
&\int_{\Phi} \dd\phi_{2} \dots \int_{\Phi} \dd\phi_{n} \Bigl| \sum_{g\in\caC_{n}} \prod_{\{i,j\} \in \caE(g)} \zeta(X_{i},X_{j}) \Bigr| \leq (\e{a k_{1}} - 1) \e{2b k_{1}}.
\end{split}
\ee
In particular, the Taylor series of the pressure $p_{\Lambda}(\bsz)$ is absolutely convergent in $D$, uniformly in $\Lambda$.

For (b) we need to control the logarithm of the derivative of $p_{\Lambda}$. It is not entirely straightforward as we need both lower and upper bounds for $\frac\partial{\partial z_{k}} p_{\Lambda}$. We have
\be
\label{ratio part fcts}
\frac{\partial p_{\Lambda}}{\partial z_{k}} = \frac1V \frac{\partial}{\partial z_{k}} \log Z_{\Lambda}(\bsz) = \frac1V \frac1{Z_{\Lambda}(\bsz)} \frac{\partial Z_{\Lambda}}{\partial z_{k}}(\bsz).
\ee
From the definition \eqref{def part fct} of the partition function, we get
\begin{multline}
\frac{\partial Z_{\Lambda}}{\partial z_{k}}(\bsz) = V \sum_{n\geq1} \frac1{(n-1)!} \sum_{k_{2},\dots,k_{n}\geq1} z_{k_{1}} \dots z_{k_{n}} \int_{\Lambda^{n-1}} \dd x_{2} \dots \dd x_{n} \int_{\Phi^{n-1}} \dd\phi_{2} \dots \dd\phi_{n} \\
\exp\Bigl\{ - \sum_{j=2}^{n} U(X,X_{j}) - \sum_{2\leq i<j\leq n} U(X_{i},X_{j}) \Bigr\}.
\end{multline}
We set $X=(k,0,0)$. The formula holds because of translation and rotation invariance, and because $\int\dd\phi_{1} = 1$. We observe that $\frac{\partial Z_{\Lambda}}{\partial z_{k}}$ is a partition function where each molecule $X_{j}$ gets the extra factor $\e{-U(X,X_{j})}$. We can again perform a cluster expansion or use the exponential formula of combinatorial structures. It is indeed convergent thanks to \eqref{strong KP}. We get
\begin{multline}
\frac{\partial Z_{\Lambda}}{\partial z_{k}}(\bsz) = V \exp \biggl\{ \sum_{n\geq1} \frac1{n!} \sum_{k_{1}, \dots, k_{n} \in \bbN} z_{k_{1}} \dots z_{k_{n}} \int_{\Lambda^{n}} \dd x_{1} \dots \dd x_{n} \int_{\Phi^{n}} \dd\phi_{1} \dots \dd\phi_{n} \\
\prod_{j=1}^{n} \e{-U(X,X_{j})} \sum_{g \in \caC_{n}} \prod_{\{i,j\}\in \mathcal{E}(g)} \zeta(X_{i},X_{j}) \biggr\}.
\end{multline}
This allows us to combine it with the cluster expansion of $Z_{\Lambda}$ in \eqref{ratio part fcts} and we get
\begin{multline}
\label{derivative pressure}
\frac{\partial p_{\Lambda}}{\partial z_{k}}(\bsz) = \exp \biggl\{ \sum_{n\geq1} \frac1{n!} \sum_{k_{1}, \dots, k_{n} \in \bbN} z_{k_{1}} \dots z_{k_{n}} \int_{\Lambda^{n}} \dd x_{1} \dots \dd x_{n} \int_{\Phi^{n}} \dd\phi_{1} \dots \dd\phi_{n} \\
\Bigl( \prod_{j=1}^{n} \bigl( 1 + \zeta(X,X_{j}) \bigr) - 1 \Bigr) \sum_{g \in \caC_{n}} \prod_{\{i,j\}\in \mathcal{E}(g)} \zeta(X_{i},X_{j}) \biggr\}.
\end{multline}
Next we use the identity
\be
\prod_{j=1}^{n} \bigl( 1 + \zeta(X,X_{j}) \bigr) - 1 = \Bigl[ \prod_{j=1}^{n-1} \bigl( 1 + \zeta(X,X_{j}) \bigr) - 1 \Bigr] \bigl( 1 + \zeta(X,X_{n}) \bigr) + \zeta(X,X_{n}).
\ee
It allows to prove by induction that
\be
\Bigl| \prod_{j=1}^{n} \bigl( 1 + \zeta(X,X_{j}) \bigr) - 1 \Bigr| \leq \e{b \sum_{j=1}^{n} k_{j}} \sum_{j=1}^{n} |\zeta(X,X_{j})|.
\ee
The integrand of \eqref{derivative pressure} is then less than
\be
\begin{split}
\sum_{k_{1} \in \bbN} |z_{k_{1}}| \e{bk_{1}} &\int_{\Lambda} \dd x_{1} \int_{\Phi} \dd\phi_{1} |\zeta(X,X_{1})| \biggl( 1 + \sum_{n\geq2} \frac1{(n-1)!} \sum_{k_{2},\dots,k_{n} \in \bbN} |z_{2} \dots z_{n}| \e{b \sum_{i=2}^{n} k_{i}} \\
&\qquad \int_{\Lambda^{n-1}} \dd x_{2} \dots \dd x_{n} \int_{\Phi^{n-1}} \dd\phi_{2} \dots \dd\phi_{n} \Bigl| \sum_{g \in \caC_{n}} \prod_{\{i,j\} \in \mathcal{E}(g)} \zeta(X_{i},X_{j}) \Bigr| \biggr) \\
&\leq \sum_{k_{1} \in \bbN} |z_{k_{1}}| \e{(a+3b) k_{1}} \int_{\Lambda} \dd x_{1}  \int_{\Phi} \dd\phi_{1} |\zeta(X,X_{1})| \\
&\leq ak.
\end{split}
\ee
We bounded the parenthesis by $\e{(a+2b) k_{1}}$ using \eqref{THE bound}. The last inequality follows from Assumption \ref{ass KP}.
\end{proof}

\medskip\noindent
{\bf Acknowledgements.} The authors are grateful to the Hausdorff Institute for making this work possible. S.~J. and D.~T. acknowledge helpful discussions with E.~Presutti. We thank the referee for many useful comments. S.~J. is supported by ERC Advanced Grant 267356 VARIS of Frank den Hollander. S.~T. and D.~U. are partially supported by EPSRC grant EP/G056390/1. D.~T. is partially supported by the FP7-REGPOT-2009-1 project
``Archimedes Center for Modeling, Analysis and Computation" (under grant agreement no 245749).

\end{document}